\documentclass[twocolumn,prc,amssymb, aps,superscriptaddress, 
amsmath]{revtex4}

\usepackage{color}
\usepackage{amsmath,mathtools,booktabs,
            microtype} %,siunitx}

\usepackage{graphicx}
\usepackage{dcolumn}
\usepackage{bm}
\usepackage{multirow}
\usepackage{times}
\usepackage{url}
\usepackage{tikz}
\usepackage[version=4]{mhchem}

\usetikzlibrary{arrows,decorations.pathmorphing}
\usetikzlibrary{arrows.meta}
\usetikzlibrary{calc}

\definecolor{violet}{HTML}{602969}
\definecolor{red}{HTML}{FC0009}
\definecolor{orange}{HTML}{FF6319}
\definecolor{green}{HTML}{00933C}
\definecolor{blue}{HTML}{0036A6}
\definecolor{yellow}{HTML}{FFBE00}
\definecolor{lightgrey}{HTML}{A7A9AC}

\newcommand{\be}{\begin{equation}}
\newcommand{\ee}{  \end{equation}}
\newcommand{\ba}{\begin{eqnarray}}
\newcommand{\ea}{  \end{eqnarray}}

\begin{document}

\title{Eigenstate thermalization hypothesis versus
  Bohigas-Giannoni-Schmit conjecture: a comparison}

\author{Hans A. \surname{Weidenm\"uller}}
\email{haw@mpi-hd.mpg.de}
\affiliation{Max-Planck-Institut f\"ur Kernphysik, Saupfercheckweg 1,
  D-69117 Heidelberg, Germany}

\date{\today}

%%%%%%%%%%%%%%%%%%%%%%%%%%%%%%%%%%%%%%%%%555

\begin{abstract}Thermalization of a closed chaotic quantum system is commonly
  addressed in terms of the ``eigenstate thermalization hypothesis''
  (ETH). An alternative approach uses the Bohigas-Giannoni-Schmit
  (BGS) conjecture. The comparison shows that the two approaches
  differ significantly. In contrast to ETH, BGS fully uses the
  statistical properies of the chaotic Hamiltonian. In both
  approaches, the criterion for thermalization is similar. BGS goes
  beyond ETH in predicting quantitatively the time dependence of
  thermalization.
\end{abstract}

%%%%%%%%%%%%%%%%%%%%%%%%%%%%%%%%%%%%%%%%%%%%%%

\maketitle

%%%%%%%%%%%%%%%%%%%%%%%%%%%%%%%%%%%%%%%%%%%%%%%%%%%%%%%%%%%%%%%%%%%%%%%%

\section{Introduction}
\label{int}

Thermalization postulates that asymptotically (time $t \to \infty$)
the expectation value ${\rm Tr} (A \rho(t))$ of a classical observable
(represented in Hilbert space by the Hermitean operator $A$) in a
closed quantum system ${\cal S}$ with density matrix $\rho(t)$ tends
toward its equilibrium value,
\ba
\label{in1}
{\rm Tr} (A \rho(t)) \to {\rm Tr} (A \rho_{\rm eq}) \ .
\ea
Thermalization is understood as the quantum analog of ergodicity in
classical statistical mechanics. Therefore, the relation~(\ref{in1})
is expected to hold in the semiclassical regime of high excitation
energy $E$ and for systems with small energy spread $\delta E \ll
E$. Reviews are given in Refs.~\cite{Bor16, Dal16, Aba19}. In recent
years, thermalization has become a topic of considerable interest in
several fields as for instance, quantum circuits~\cite{A}, quantum
lattice systems~\cite{B}, open quantum systems~\cite{C}, and
low-entanglement states~\cite{D}.

The asymptotic relation~(\ref{in1}) cannot be expected to hold for
every quantum system. For the system $S$ with Hamiltonian $H$, that is
seen by writing ${\rm Tr}(A \rho(t))$ in the eigenstate representation
of $H$ with eigenstates $\alpha, \beta, \ldots$ and eigenvalues
$E_\alpha, E_\beta, \ldots$ where
\ba
\label{in2}
{\rm Tr} ( A \rho(t)) = \sum_{\alpha \beta} A_{\alpha \beta} \big( \exp \{
- i E_\beta t \} \Pi_{\beta \alpha} \exp \{ + i E_\alpha t \} \big)\ .
\ea
We put $\hbar = 1$ throughout. The terms in round brackets on the
right-hand side represent the elements of the density matrix. The
time-independent statistical operator $\Pi$ has unit trace and defines
the occupation amplitudes of the system ${\cal S}$ in Hilbert
space. The diagonal elements $\rho_{\alpha \alpha}$ of $\rho(t)$ are
independent of time, and the non-diagonal elements $\rho_{\beta
  \alpha}$ oscillate periodically with frequency $(E_\alpha -
E_\beta)$. Thus, ${\rm Tr} ( A \rho(t))$ is an oscillatory function of
time, and thermalization does not occur in general. Indeed, integrable
systems and systems showing many-body localization do not
thermalize~\cite{Dal16, Aba19}. The study of thermalization has been
focused on quantum systems that are chaotic in the classical
limit. The eigenfunctions and eigenvalues of such systems display
statistical properties. Specifically, for a time-reversal-invariant
chaotic quantum system (which we consider in what follows), the
Bohigas-Giannoni-Schmit (BGS) conjecture states~\cite{Boh84} that the
spectral fluctuations of eigenvalues and eigenfunctions are the same
as for the GOE, the Gaussian Orthogonal Ensemble of Random
Matrices~\cite{Meh04}. Thermalization is a consequence of and follows
from averaging over these statistical fluctuations.

In a seminal paper, M. Srednicki~\cite{Sre99} used the ``eigenstate
thermalization hypothesis'' to investigate thermalization in a closed
chaotic quantum system. The statistical description of chaotic
many-body quantum systems based on the Bohigas-Giannoni-Schmit
conjecture~\cite{Boh84} developed in Refs.~\cite{Wei24, Wei25} offers
an alternative access to thermalization. In the present paper, we
compare the two approaches.

\section{Eigenstate thermalization hypothesis}
\label{eig}

In Ref.~\cite{Sre99}, the statistical operator $\Pi$ in
Eq.~(\ref{in2}) is defined in terms of a single normalized state
$|\psi(t)\rangle$. In the eigenstate representation $|\psi(t)\rangle$
is written as
\ba
\label{e1}
| \psi(t) \rangle = \sum_\alpha c_\alpha \exp \{ - i E_\alpha t \} |
\alpha \rangle
\ea
where
\ba
\label{e2}
\sum_\alpha | c_\alpha |^2 = 1 \ .
\ea
Density matrix and statistical operator are, respectively, given by
\ba
\label{e3}
\rho(t) = | \psi (t) \rangle  \langle \psi(t) | \ , \ \Pi_{\alpha \beta}
= c_\alpha c_\beta^* 
\ea
so that
\ba
\label{e4}
{\rm Tr} (A \rho(t)) = \sum_{\alpha \beta} c_\alpha c^*_\beta \exp \{ i
(E_\beta - E_\alpha) t \} A_{\alpha \beta} \ .
\ea
The elements $A_{\alpha \beta}$ of the operator $A$ possess statistical
properties and are written as~\cite{Sre99}
\ba
\label{e5}
A_{\alpha \beta} = {\cal A}(E) \delta_{\alpha \beta} + \exp \{ - S / 2 \}
  f(E, \omega) R_{\alpha \beta} \ .
\ea
Here ${\cal A}(E)$ and $f(E, \omega)$ are smooth real functions of $E
= (1/2) (E_\alpha + E_\beta)$ and $\omega = E_\alpha - E_\beta$,
$S(E)$ is the thermodynamic entropy of the system, and $R_{\alpha
  \beta} = R_{\beta \alpha}$ is a zero-centered real Gaussian random
variable with variance
\ba
\label{e6}
\langle R_{\alpha \beta} R_{\gamma \delta} \rangle = \delta_{\alpha
  \gamma} \delta_{\beta \delta} + \delta_{\alpha \delta} \delta_{\beta
  \gamma} \ .
\ea
Here and in what follows, the angular brackets denote the ensemble
average.

The function ${\cal A}(E)$ is the average of the diagonal elements of
$A$. The average is taken over a large energy interval centered on
energy $E$. It is argued~\cite{Sre99} that for $E$ in the
semiclassical regime, ${\cal A}(E)$ is equal to the thermal average
$\langle A \rangle_T$ of $A$. The function $f(E, \omega)$ describes
the average dependence of the non-diagonal matrix elements of $A$ on
$E$ and $\omega$. The factor $\exp \{ - S / 2 \}$ accounts for the
smallness of these elements in the semiclassical regime. The factor
$R_{\alpha \beta}$ accounts for their stochastic features.

In Ref.~\cite{Sre99}, the expansion coefficients $c_\alpha, c^*_\beta$
and the energies $E_\alpha$, $E_\beta$ in Eq.~(\ref{e4}) are taken to
be nonstatistical. Insertion of Eq.~(\ref{e5}) into Eq.~(\ref{e4}) and
averaging over the distribution yields with Eq.~(\ref{e2}) the result
\ba
\label{e7}
\big\langle {\rm Tr} (A \rho(t) ) \big\rangle = {\cal A}(E) \ .
\ea
The time dependence of ${\rm Tr} (A \rho(t))$ disappears after
averaging. The variance of ${\rm Tr} (A \rho(t))$ is similarly found
to be
\ba
\label{e8}
&& \bigg\langle \bigg( {\rm Tr} (A \rho(t) ) - {\cal A}(E) \bigg)^2
\bigg\rangle = \exp \{ - S \} \\
&& \times \sum_{\alpha \beta} \bigg[ f^2(E, \omega) \bigg( \Pi_{\alpha \beta}
\Pi_{\beta \alpha} \nonumber \\
&& \ \ \ + (\exp \{ i (E_\alpha - E_\beta) t \} \Pi_{\alpha \beta})^2) \bigg)
\bigg] \ . 
\nonumber
\ea
Since $\omega = E_\alpha - E_\beta$, the sum over $(\alpha, \beta)$
extends also over $f^2(E, \omega)$. Because of the factor $\exp \{ - S
\}$, the variance of ${\rm Tr} (A \rho(t))$ and with it, the
fluctuations of ${\rm Tr} (A \rho(t))$ about its average are
negligibly small in the semiclassical regime~\cite{Sre99}.

The time dependence of ${\rm Tr} (A \rho(t))$ is determined as
follows. For a general system with statistical operator $\Pi$, the
average $\langle E \rangle$ and the quantum uncertainty $\Delta_S$ of
the energy $E$ of the system $S$ are given by
\ba
\label{e9}
\langle E \rangle &=& \sum_\alpha E_\alpha \Pi_{\alpha \alpha} \ , \nonumber \\
\Delta_S^2 &=& \sum_\alpha (E_\alpha - \langle E \rangle)^2 \Pi_{\alpha \alpha} \ .
\ea
In Ref.~\cite{Sre99} it is shown that the difference of the thermal
averages
\ba
\label{e10}
\langle A^2 \rangle_T - (\langle A \rangle_T)^2 
\ea
is of order $\Delta_S^2$. With $\langle A \rangle_T = {\cal A}$, that
fact accounts for finite-size fluctuations of ${\rm Tr} (A \rho(t))$
about its mean value ${\cal A}$. The fluctuations are shown to decay
in time. As a consequence, ${\rm Tr} (A \rho(t))$ approaches ${\rm Tr}
(A \rho_{\rm eq}) = \langle A \rangle_T$ for long times. The
difference ${\rm Tr} (A \rho(t)) - {\rm Tr} (A \rho_{\rm eq})$ is
shown to be proportional to the Fourier transform $C(t)$ of $| f(E,
\omega) |^2$ with respect to $\omega$. The rate of change of $C(t)$
with time is determined by the band width in $\omega$ of $f(E,
\omega)$. Since $| f(E, \omega) |^2 \geq 0$, the time-dependent
function $C(t)$ oscillates about zero. Maxima and minima of the
oscillations decrease monotonically with time. The analytical
dependence of $| f(E, \omega) |^2$ on $\omega$ and that of $C(t)$ on
time $t$ are not given explicitly in Ref.~\cite{Sre99} nor is it
indicated how these might be determined from the Hamiltonian of the
system.

\section{Bohigas-Giannoni-Schmit (BGS) conjecture}
\label{boh}

In Ref.~\cite{Wei24}, the BGS conjecture has been applied to a chaotic
many-body quantum system. We here repeat the arguments in a nutshell
and focus attention on the results. For detailed justification, we
refer to Ref.~\cite{Wei24} and, for explicit calculations, to
Ref.~\cite{Wei25}.

The Hamiltonian $H = H_{\rm HF} + V$ is written as the sum of an
(integrable) Hartree-Fock (HF) Hamiltonian $H_{\rm HF}$ and a residual
interaction $V$. The HF eigenstates are labeled $m, n, \ldots$, the HF
energies are written as ${\cal E}_m, {\cal E}_n, \ldots$. The residual
interaction $V$ mixes the HF eigenstates. Diagonalization of $H$ by an
orthogonal matrix with elements $O_{m \alpha}$ yields the eigenvalues
$E_\alpha$. In the HF basis, the diagonalized Hamiltonian is
\ba
\label{b1}
H_{m n} = \sum_\alpha O_{m \alpha} E_\alpha O_{n \alpha} \ .
\ea
The system is assumed to be chaotic in the classical limit. In the
quantum case, the BGS conjecture implies that $V$ mixes the HF states
locally sufficiently strongly to cause the eigenvalues $E_\alpha$ and
the elements $O_{m \alpha}$ of the diagonalizing orthogonal matrix
to obey GOE statistics. They do so at any energy $E$ within an energy
interval of size $\Delta$ centered on $E$. The system becomes
analytically tractable by introducing an ensemble of random
Hamiltonians that share with $H$ these statistical properties. The
ensemble is obtained by promoting the eigenvalues $E_\alpha$ and the
elements $O_{m \alpha}$ of the orthogonalizing matrix to random
variables. Within the interval $\Delta$, the eigenvalues $E_\alpha$
obey Wigner-Dyson statistics.  The elements $O_{m \alpha}$ are
zero-centered Gaussian-distributed random variables with second
moments
\ba
\label{b2}
\big\langle O_{m \alpha} O_{n \beta} \big\rangle = \delta_{m n}
\delta_{\alpha \beta} F(\overline{E}_\alpha - {\cal E}_m)
\ea
where
\ba
\label{b3}
F(\overline{E}_\alpha - {\cal E}_m) &=& \frac{1}{\sqrt{2 \pi}
  \rho(\overline{E}_\alpha) \Delta} \nonumber \\
&& \times \exp \{ - (\overline{E}_\alpha - {\cal E}_m)^2 / (2 \Delta^2) \} \ .
\ea
Here $\rho(E)$ is the average level density, and $\overline{E}_\alpha$
is the ensemble average of $E_\alpha$. The statistical properties of
the GOE are derived~\cite{Meh04} in the limit of infinite matrix
dimension. Full agreement with the statistical properties of the GOE
is, therefore, only attained if the number
\ba
\label{b4}
N = \rho(E) \Delta
\ea
of eigenvalues in the interval $\Delta$ obeys $N \gg 1$. That
condition holds in the semiclassical regime.

The function ${\rm Tr} (A \rho(t))$ is written as
\ba
\label{b5}
{\rm Tr} (A \rho(t)) &=& \sum_{m n r s \alpha \beta} (O_{m \beta} A_{m n}
O_{n \alpha}) \exp \{ - i E_\alpha t \} \nonumber \\
&& \ \ \ \times (O_{r \alpha} \Pi_{r s} O_{s \beta}) \exp \{ + i E_\beta t \} \ .        
\ea
Eq.~(\ref{b5}) displays explicitly the dependence of ${\rm Tr} (A
\rho(t))$ on the random variables $O_{m \alpha}$ and $E_\alpha$. The
function ${\rm Tr} (A \rho(t))$is a stochastic process. The average
and the variance of ${\rm Tr} (A \rho(t))$ are calculated in an
asymptotic expansion in inverse powers of $N$ where only terms of
leading order are kept.

For the average of ${\rm Tr} (A \rho(t))$ the result is~\cite{Wei24}
\ba
\label{b6}
&&\big\langle {\rm Tr} (A \rho(t)) \big\rangle = \sum_k \frac{p_k}
{\sqrt{2} \pi \rho_k \Delta} {\rm Tr}_k (A) \\
&& + {\rm Tr} \big( A \exp \{ - i H_{\rm HF} t \} \Pi \exp \{ + i H_{\rm HF}
t \} \big) \exp \{ - \Delta^2 t^2 \} \ . \nonumber
\ea
For the first term on the right-hand side, the total spectrum has been
divided into intervals labeled $k$ of width $\Delta$ each, with
$\rho_k$ the average level density in the interval $k$, the trace
${\rm Tr}_k$ extending only over the states within the interval $k$,
and $p_k$ an approximation to the partial trace of $\Pi$ extended over
the HF eigenstates within the interval $k$.

In Ref.~\cite{Wei25} it is shown that the correlation function
\ba
\label{b7}
\big\langle {\rm Tr} (A \rho(t_1)) {\rm Tr} (A \rho(t_2)) \big\rangle
- \big\langle {\rm Tr} (A \rho(t_1)) \big\rangle \big\langle {\rm Tr}
(A \rho(t_2)) \big\rangle
\ea
vanishes for $N \to \infty$. It follows that the result~(\ref{b6})
holds for almost all members of the random-matrix ensemble in
Eq.~(\ref{b1}), with the exception of a set of measure zero. The
measure is the integration measure for the random variables $O_{m
  \alpha}$ and $E_\alpha$. We use the result~(\ref{b7}) to omit in
Eq.~(\ref{b6}) the angular brackets.

Eq.~(\ref{b6}) shows that thermalization as defined in Eq.~(\ref{in1})
does not hold in general. Thermalization occurs only if the
statistical operator $\Pi$ is confined to an energy interval bounded
from above by $\Delta$ so that the quantum uncertainty $\Delta_S$
defined in Eq.~(\ref{e9}) is less than $\Delta$. Then only a single
term (with index $k_0$, say) contributes to the sum on the right-hand
side of Eq.~(\ref{b6}), we have $p_{k_0} = 1$, and $(1 / \sqrt{2} \pi
\rho_{k_0} \Delta) {\rm Tr}_{k_0} (A)$ is equal to the equilibrium
value ${\rm Tr} (A \rho_{\rm eq})$ because the Boltzmann factor is
approximately constant within the energy interval $\Delta$ and cancels
in numerator and denominator. Thus, for $\Delta_S < \Delta$,
Eq.~(\ref{b6}) takes the form
\ba
\label{b8}
&& {\rm Tr} (A \rho(t)) = {\rm Tr} (A \rho_{\rm eq}) \\
&& + {\rm Tr} \big( A \exp \{ - i H_{\rm HF} t \} \Pi
\exp \{ + i H_{\rm HF} t \} \big) \exp \{ - \Delta^2 t^2 \} \ .
\nonumber
\ea
The condition $\Delta_S < \Delta$ quantifies the condition $\delta E
\ll E$ mentioned in the Introduction.

\section{Comparison. Discussion. Conclusions}

In comparing results we refer to Ref.~\cite{Sre99} with the letters
ETH and to Ref.~\cite{Wei24} with the letters BGS. Both ETH and BGS
aim at a realistic description of the thermalization process in a
chaotic system. Therefore, it is meaningful to compare the results of
both approaches term by term. Comparing Eqs.~(\ref{e4}) and
(\ref{b5}), we identify
\ba 
\label{c1}
A_{\alpha \beta} &\leftrightarrow& \sum_{m n} (O_{m \beta} A_{m n} O_{n \alpha})
\ , \nonumber \\
c_\alpha c^*_\beta &\leftrightarrow& \sum_{r s} (O_{r \alpha} \Pi_{r s}
O_{s \beta}) \ .
\ea
The first identification is obvious. The second identification is
valid because when $\Pi$ has rank one we have $\Pi_{r s} = \pi_r
\pi^*_s$ which yields $c_\alpha = O_{r \alpha} \pi_r$, $c^*_\beta =
O_{s \beta} \pi^*_s$. We also identify
\ba
\label{c3}
\exp \{ - S \} \leftrightarrow 1 / N \ . 
\ea
That identification is consistent with the physical significance of
$\exp \{ - S \}$ and with the fact that in both approaches only terms
of leading order in these small parameters are kept.

To compare the eigenstate thermalization hypothesis for $A_{\alpha
  \beta}$ in Eq.~(\ref{e5}) with the statistical properties of
$\sum_{m n} (O_{m \alpha} A_{m n} O_{n \beta})$ in Eq.~(\ref{c1}) we
calculate mean value and variance of $\sum_{m n} (O_{m \alpha} A_{m n}
O_{n \beta})$. The average is given by
\ba
\label{c2}
\big\langle \sum_{m n} (O_{m \alpha} A_{m n} O_{n \beta}) \big\rangle = 
\frac{\delta_{\alpha \beta}} {\sqrt{2 \pi} \rho(\overline{E}_\alpha) \Delta}
\qquad \qquad \\ \nonumber
\qquad \qquad \times \sum_m A_{m m} \exp \{ - (\overline{E}_\alpha -
       {\cal E}_m)^2 / (2 \Delta^2 \} \ .
\ea
The variance is given in the Appendix. It is small of order $1 / N$.

The expressions for $\langle A_{\alpha \beta} \rangle$ in ETH and BGS
are similar. Both are diagonal in $\alpha, \beta$, and are given in
terms of energy averages. In ETH, the average ${\cal A}$ of $A$ is, in
the semiclassical regime, given by the thermal average $\langle A
\rangle_T$. In BGS, $\langle A_{\alpha \beta} \rangle$ is the
normalized trace of $A$ taken over an energy interval of width
$\Delta$ centered on $\overline{E}_\alpha$. In BGS, a thermal average
cannot appear without explicit reference to the density matrix of the
system. The last term in Eq.~(\ref{e5}) and the terms on the
right-hand side of Eq.~(\ref{a1}) are small of first order in the
parameters~(\ref{c3}). BGS goes beyond ETH because Eq.~(\ref{a1})
yields an explicit expression for the square of the function $f(E,
\omega)$ in Eq.~(\ref{e5}).

For $\langle {\rm Tr} (A \rho(t) \rangle$, both ETH and BGS yield a
time-independent term, see Eqs.~(\ref{e7}) and (\ref{b8}). These terms
describe the asymptotic value of $\langle {\rm Tr} (A \rho(t))
\rangle$ attained for large times. In ETH, the term is equal to ${\cal
  A}(E)$ in Eq.~(\ref{e5}) and, in the semiclassical regime, to
$\langle A \rangle_T$. In BGS, the term is equal to ${\rm Tr} (A
\rho_{\rm eq})$ provided that $\Delta_S < \Delta$. The two results are
basically equivalent but BGS states more explicitly the conditions for
thermalization to occur.

Beyond the similarity shown by that comparison, the
relations~(\ref{c1}) actually point to a striking difference between
ETH and BGS. In both approaches, the operator $A$, written in the
eigenbasis of $H$, acquires statistical properties. However, in ETH
the term $c_\alpha c^*_\beta$ is considered nonstochastic while in BGS
the expression $\sum_{r s} (O_{r \alpha} \Pi_{r s} O_{s \beta})$ is
taken as a stochastic variable in the same sense as $\sum_{m n} (O_{m
  \beta} A_{m n} O_{n \alpha})$. Consistency requires that both these
expressions are treated on the same footing. The notation $c_\alpha
c_\beta^*$ used for the matrix elements of the statistical operator
blurs that need in ETH. The difference in statistical assumptions
strongly affects the time dependence of ${\rm Tr} (A \rho(t))$.

In BGS, the time-dependent term on the right-hand side of
Eq.~(\ref{b8}) is due to the correlation of the stochastic variables
$\sum_{m n} (O_{m \beta} A_{m n} O_{n \alpha})$ and $\sum_{r s} (O_{r
  \alpha} \Pi_{r s} O_{s \beta})$ in Eq.~(\ref{b5}). The term is the
product of a factor the time dependence of which is determined by the
HF Hamiltonian, and of the factor $\exp \{ - \Delta^2 t^2 / 2 \}$, the
Fourier transform of the function $F$ in Eq.~(\ref{b3}). The scale of
the time dependence is given by $\Delta$, the correlation width of the
BGS conjecture.

The analog of the time-dependent term in Eq.~(\ref{b6}) is absent in
Eq.~(\ref{e7}). That is because the term $c_\alpha c^*_\beta$ is
asumed to be non-statistical. As described below Eq.~(\ref{e10}),
deviations from the mean value~(\ref{e7}) are ascribed in ETH to the
fluctuations of $A$ around its mean value $\langle A \rangle_T$. The
resulting non-monotonic time dependce of $C(t)$ differs qualitatively
from the Gaussian form in Eq.~(\ref{b8}).

In both ETH and BGS, ${\rm Tr} (A \rho(t))$ is a stochastic process.
In both approaches, the variance of ${\rm Tr} (A \rho(t))$ vanishes in
the limit where the small parameters~(\ref{c3}) tend to zero. From a
systematic point of view, ${\rm Tr} (A \rho(t))$ should, therefore, be
determined entirely by its mean value $\langle {\rm Tr} (A \rho(t))
\rangle$. That desideratum is met in BGS. It is not met in ETH because
the product $c_\alpha c^*_\beta$ is considered nonstatistical. 

It must be borne in mind that BGS is obtained in the framework of the
stochastic many-body Hamiltonian in Eqs.~(\ref{b1}, \ref{b2}) and
(\ref{b3}) while ETH addresses chaotic quantum systems in full
generality. Our comparison, therefore, applies strictly only to
many-body systems. However, the problems of ETH displayed by the
comparison are obviously generic. Moreover, the parametrization in
Eqs.~(\ref{b1}, \ref{b2}) and (\ref{b3}) of the stochastic Hamiltonian
is not as restrictive as it may seem, for the following reason. The
parametrization uses the Hartree-Fock approach. The integrable HF
Hamiltonian provides a scaffolding for the spectrum of the chaotic
Hamiltonian $H$. The residual interaction $V$ locally mixes the HF
eigenstates and leads to GOE fluctuations within the correlation width
$\Delta$. For that to happen, it is not necessary to invoke
statistical assumptions on $V$. That approach can be generalized. It
is often the case that an integrable system with Hamiltonian $H_{\rm
  int}$ becomes chaotic upon adding a perturbation $\tilde{V}$ to
$H_{\rm int}$. The integrable system then serves as a scaffolding for
the spectrum. The perturbation $\tilde{V}$ mixes the eigenstates of
$H_{\rm int}$ locally in the same way as does $V$ for the HF
Hamiltonian $H_{\rm HF}$. That fact, combined with the BGS conjecture,
then leads to a parametrization of $H_{\rm int} + \tilde{V}$ similar
to that of Eqs.~(\ref{b1}) to (\ref{b3}). If, in addition, the level
density of the system increases strongly with excitation energy, one
obtains a result of the form of Eq.~(\ref{b8}). That is why we believe
that Eq.~(\ref{b8}) is valid for a class of chaotic quantum systems
that encompasses but is wider than the class of chaotic many-body
systems. In all these cases, BGS replaces the semiquantitative ETH
estimate for the time dependence of ${\rm Tr} (A \rho(t))$ by an
analytic result. The time scale is given by the inverse of the
correlation width $\Delta$ of the BGS conjecture.

It seems plausible that our results apply in an even wider context
because the BGS conjecture applies to chaotic quantum systems in
general. The difficulty in extending our argument to a more general
situation is technical. To formulate the BGS conjecture
quantitatively, we need the scaffolding provided by the integrable
Hamiltonian and the reference to its eigenvalues ${\cal E}_m$. That
leads to Eq.~(\ref{b2}). So far we have seen no way of bypassing that
construction.

To sum up: ETH and BGS differ signifcantly. ETH uses a well-justified
ad-hoc parametrization of the statistical matrix $A_{\alpha \beta}$
but disregards statistical properties of the term $c_\alpha
c^*_\beta$. BGS fully uses the statistical properties of the
Hamiltonian implied by the BGS conjecture. In contrast to ETH, BGS
predicts quantitatively the time dependence of ${\rm Tr} ( A
\rho(t))$.

The author declares no conflicts of interest regarding this
paper. There are no data beyond the ones presented in the paper. There
was no outside funding.

\section*{Appendix}

The variance of $\langle O_{m \alpha} A_{m n} O_{n \beta} \rangle$ is
obtained by calculating the correlated part of the square of that
expression. For simplicity we assume that the matrix $A_{m n}$ is real
and symmetric. We obtain
\ba
\label{a1}
&& \langle (O_{m \alpha} A_{m n} O_{n \beta})^2 \rangle_{\rm corr} = \frac{1}
{2 \pi \Delta^2 \rho(\overline{E}_\alpha) \rho(\overline{E}_\beta)}
\\ \nonumber
&& \qquad \qquad \times {\rm Tr} \bigg( A \exp \{ (\overline{E}_\alpha
- H_{\rm HF})^2 / (2 \Delta^2) \} \\ \nonumber
&& \qquad \qquad \qquad \times A \exp \{ (\overline{E}_\beta -
H_{\rm HF})^2 / (2 \Delta^2) \bigg) \ .
\ea
The right-hand side of that equation is the product of the normalized
trace and of a factor $\propto 1 / N$ and is, therefore, small of
order $1 / N$.

%%%%%%%%%%%%%%%%%%%%%%%%%%%%%%%%%

\end{document}